\documentclass[reprint,twocolumn,prb,amsmath,amssymb,aps,nofootinbib,superscriptaddress]{revtex4-2} 

\usepackage{graphicx}  
\usepackage{dcolumn}   
\usepackage{bm}        
\usepackage{hyperref}  
\hypersetup{colorlinks=true, linkcolor=blue!95!black!85!yellow, citecolor=blue!95!black!85!yellow, urlcolor=blue!95!black!85!yellow} 
\usepackage{physics}   
\usepackage{newtxtext} 
\usepackage[cmintegrals]{newtxmath} 
\usepackage{booktabs}  
\usepackage{multirow}  
\usepackage{xcolor}    
\usepackage{float}     
\usepackage{xcolor}

\usepackage{tikz}

\preprint{APS/123-QED}

\begin{document}
	
	
	\title{Pressure-driven structural phase transition unlocks multifunctionality in KMgX (X = P, As, Sb, and Bi) compounds: A first-principles study}
	
	\author{Chetna Saini}
    \affiliation{Department of Chemistry, Indian Institute of Technology Roorkee, Roorkee-247667, Uttarakhand, India}
    
    \author{Neha Sadanandan}
    \affiliation{Department of Chemistry, Indian Institute of Technology Roorkee, Roorkee-247667, Uttarakhand, India}
    
	\author{Tashi~Nautiyal}
	\affiliation{Department of Physics, Indian Institute of Technology Roorkee, Roorkee-247667, Uttarakhand, India}
	
	\author{Hongbin Zhang}
	\affiliation{Institute of Materials Science, Technology University of Darmstadt, 64287 Darmstadt, Germany}

	\author{Vikrant~Chaudhary} \email{Corresponding author: chaudhary@tmm.tu-darmstadt.de}
    \affiliation{Institute of Materials Science, Technology University of Darmstadt, 64287 Darmstadt, Germany} 
    \affiliation{Physics Department and CSMB, Humboldt-Universit\"at zu Berlin, 12489 Berlin, Germany}
    
	\author{Hem C. Kandpal} \email{Corresponding author: hem.kandpal@cy.iitr.ac.in}
    \affiliation{Department of Chemistry, Indian Institute of Technology Roorkee, Roorkee-247667, Uttarakhand, India} 
\date{\today}
	
\begin{abstract}
		The search for materials with multifunctional properties has attracted significant attention due to their potential applications in various energy-related devices. Pressure-induced phase transitions provide an effective strategy for accessing different structural phases of a material without altering its chemical composition, thereby enabling the tuning of its physical properties and expanding its functional applications. In this work, we investigate the previously unexplored orthorhombic ($Pnma$) phase of the KMgX (X = P, As, Sb, and Bi) family using first-principles calculations and identify a pressure-induced structural transition from a tetragonal to orthorhombic phase. The stability of the pressure-accessible orthorhombic structure is rigorously confirmed by equation of state analysis together with phonon, elastic, and formation-enthalpy calculations, establishing its viability for further investigation. Optical properties calculated within the $G_0W_0$--Bethe--Salpeter equation (BSE) framework, incorporating quasiparticle corrections and excitonic effects, exhibit direct dipole-allowed transitions at the $\Gamma$ point and strong visible-light absorption with coefficients approaching $10^5~\mathrm{cm}^{-1}$. Consequently, KMgAs and KMgSb achieve spectroscopic limited maximum efficiencies (SLME) of $27.12\%$ and $26.40\%$, respectively, at a thin-film thickness of $0.6~\mu\mathrm{m}$. Furthermore, thermoelectric transport calculations predict (zT) values of 0.65 and 0.58 at $900~\mathrm{K}$ for p-type and n-type KMgSb, respectively, demonstrating its potential for both legs of thermoelectric devices. These values are likely conservative, as the Slack model tends to overestimate thermal conductivity. Overall, the pressure-accessible orthorhombic phase of the KMgX family emerges as a stable multifunctional semiconductor with coupled photovoltaic and thermoelectric energy-conversion capabilities.
\end{abstract}
	
\maketitle
	
\section{Introduction} 
\setlength{\parindent}{3em}
Materials form the backbone of modern device fabrication and their practical applications \cite{Gutowski17,Evans2016}. The discovery of new materials is a challenging and time-consuming process that requires substantial effort, including theoretical prediction, computational screening, extensive experimental validation, and numerous trial and error investigations. Consequently, alternative strategies for discovering functional materials have gained increasing attention. Among these, engineering existing materials into different structural phases through external stimuli, such as pressure and temperature, has emerged as an effective approach for tailoring their physical and chemical properties without altering their chemical composition \cite{Xu2021, Zeidler17}.

The phase engineering strategy enables the realization of multifunctional materials with diverse properties, thereby expanding their potential for a broad range of technological and energy-related applications. Extensive research in this area has resulted in the identification of several important material families, including perovskites \cite{olasoji2025metal,li2017chemically}, graphene-based materials \cite{yu2017graphene,wu2023graphene,han2023orbital}, Heusler compounds \cite{zeier2016engineering,chadov2010tunable,wollmann2017heusler}, transition metal dichalcogenides \cite{manzeli20172d,ji2026large}, Zintl phase \cite{chaudhary2023first,dhawan2023electronic,hu2026manipulation}, and others, which demonstrate remarkable multifunctionality.

One cost-effective intermetallic compound, FeSb$_2$, known for exhibiting a colossal Seebeck coefficient at low temperatures \cite{Bentien07, sun10}, making it an attractive candidate for thermoelectric applications \cite{gujjar24}. Beyond thermoelectricity, FeSb$_2$ has also demonstrated remarkable potential in environmental and energy-related applications. It has been reported as an efficient adsorbent for the removal of Pb(II) ions from aqueous solutions \cite{gujjar23}, an effective adsorbent co-catalyst for the degradation of congo red dye, and a self-supported electro-catalyst for hydrogen evolution reactions (HER) \cite{dgujjar23}. These diverse functionalities highlight the versatility of FeSb$_2$ and underscore its potential as a multifunctional material for next-generation energy and environmental technologies.

Over the last decade, ternary KMgX (X = P, As, Sb, and Bi) compounds have attracted considerable attention owing to their ability to crystallize in multiple structural phases, including hexagonal \cite{bennett2012hexagonal}, tetragonal \cite{vogel1979neue}, cubic \cite{arif2016elastic}, and orthorhombic \cite{bennett2013orthorhombic} phase. Each crystallographic phase exhibits distinct physical and chemical properties, making these compounds promising candidates for a wide range of applications, particularly in thermoelectric \cite{chaudhary2023effect,lv2025unconventional} and optoelectronics \cite{chand2025enhanced}. In the cubic phase, the KMgX (X = P, As, Sb, Bi) compounds have a C1$_b$-type structure with space group $\textit{F$\bar{4}$3m}$. To the best of our knowledge, no experimental study has confirmed the existence of this structure. Despite this, theoretical studies based on DFT predict that the cubic phase of KMgX compounds has a direct band gap and are suitable for thermoelectric and photovoltaic applications \cite{chand2025enhanced,lv2025unconventional,arif2016elastic}.

Experimentally, the KMgX (X = P, As, Sb, Bi) compounds crystallize in the tetragonal phase with space group $\textit{P4/nmm}$ (No. 129) and have a Matlockite structure \cite{vogel1979neue}. Among the KMgX family, the tetragonal phase of KMgSb and KMgBi is well studied for thermoelectrics \cite{chaudhary2023effect}. Also, the tetragonal phase of KMgBi has been reported as a topological semi-metal with a type-$\mathrm{I}$ Dirac point \cite{le2017three}, and in another study, reported as a narrow band gap semiconductor \cite{zhang2017narrow}. This tetragonal phase of KMgBi is a unique member of the KMgX family, exhibiting a synergistic combination of a large ordinary Nernst effect and axis-dependent conduction polarity, which enhances its transverse thermoelectric performance \cite{ochs2024synergizing}. All members of the KMgX compound family have also been investigated for photovoltaic applications in their tetragonal phase \cite{mouchou2025thermoelectric,choudhary2020joint}.

The hexagonal and orthorhombic phases of KMgSb and KMgBi possess ferroelectric and antiferroelectric structures, respectively, suggesting their potential for applications such as high-energy-storage capacitors and electrocaloric refrigeration \cite{bennett2013orthorhombic,bennett2012hexagonal}. However, these possible applications remain largely unexplored. For all the members of the KMgX family, the hexagonal and orthorhombic phases also remain unexplored for thermoelectric and optoelectronic applications. To the best of our knowledge, no systematic investigation has been conducted to understand the thermodynamic conditions governing the stability of discussed phases or the phase transformations within the KMgX family.

Therefore, in this work, we address two important challenges: first, how one can transform one phase into another by applying pressure, and second, the feasibility, stability, and existence conditions of the orthorhombic phase. After establishing a clear understanding of the accessibility of these phases, we further investigate the orthorhombic phase for potential thermoelectric and optoelectronic applications. Thus, this work provides a comprehensive investigation of the structural stability, pressure-induced phase transitions, electronic structure, lattice dynamics, thermoelectric- and photovoltaic performance of the previously unexplored orthorhombic KMgX family, thereby opening a new pathway toward multifunctional ternary materials.
\begin{figure*}
			\centering
			\includegraphics[width=0.8\textwidth]{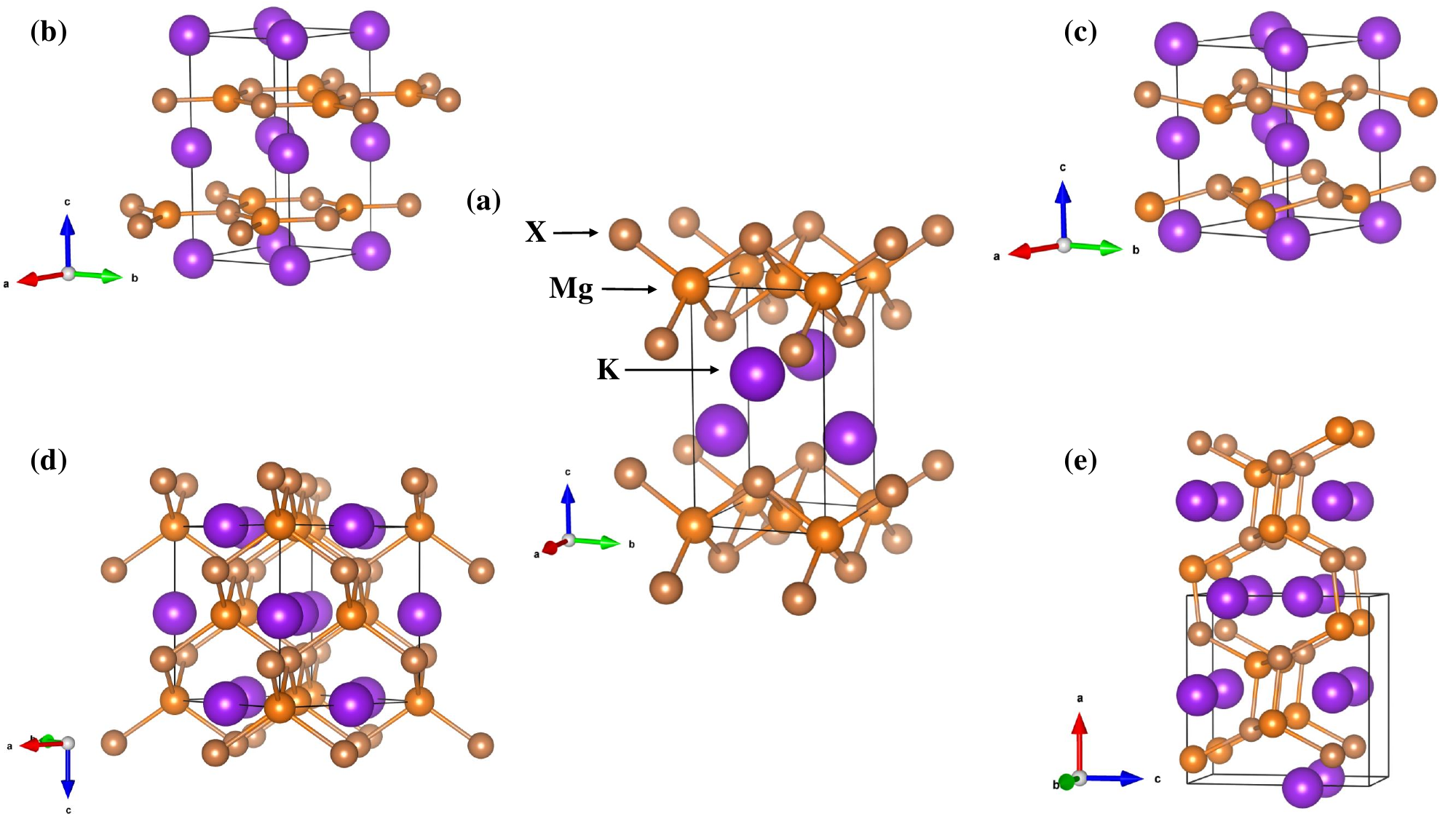}
			\caption{Crystal structures of different phases of KMgX (X = P, As, Sb, and Bi) in (a) $P4/nmm$, (b) $P6_{3}/mmc$, (c) $P6_{3}mc$, (d) $F\bar{4}3m$, and (e) $Pnma$ space groups, where purple, orange and brown colour indicate K, Mg and X, respectively.  (a), (b), and (c) have a pseudo 2-dimensional layered structure in which the K atoms are sandwiched between Mg-X layers, whereas (d) and (e) adopt a 3-dimensional network structure, with the K atoms stuffed within the Mg-X framework.}
			\label{fig:str}
\end{figure*}

\section{Computational Detail} 
\setlength{\parindent}{3em}
All first-principles calculations in the work were performed using the Vienna Ab initio Simulation Package (VASP) \cite{kresse1993ab, kresse1994ab, kresse1996efficiency, kresse1996efficient}, based on the projector augmented wave (PAW) \cite{kresse1999ultrasoft, blochl1994projector} method within a plane-wave basis set. The plane-wave basis was truncated at a kinetic energy cutoff of 550~eV, ensuring a reliable convergence of total energy and derived properties. Structural relaxations were carried out using this dense k-mesh of $10 \times 18 \times 10$, with atomic forces converged to 0.001~eV/\AA\ and total energy convergence set to $10^{-8}$~eV. 
	
 For structural optimization and preliminary electronic properties, the generalized gradient approximation (GGA) in the form of the Perdew–Burke–Ernzerhof (PBE) \cite{perdew1996generalized} functional was employed. However, since GGA functionals are known to underestimate the band gap of semiconductors, more accurate electronic properties were obtained using the Heyd–Scuseria–Ernzerhof hybrid functional (HSE06) \cite{krukau2006influence}. For accounting the relativistic effects in the KMgX family containing heavy elements such as Sb and Bi, spin–orbit coupling (SOC) was explicitly included in the electronic structure calculations.
	
The lattice dynamic properties, including the phonon band structure and Gr\"{u}neisen parameters, were evaluated using the finite displacement method within the harmonic approximation implemented in the PHONOPY \cite{togo2015first} package. For Gr\"{u}neisen parameter calculations, the volume of the unit cell was changed by  $+0.5\%$ and $-0.5\%$. Furthermore, the dielectric tensor and mechanical properties were calculated using the Density Functional Perturbation Theory (DFPT) module of VASP, and the results were used to evaluate the scattering rates. A $2 \times 3 \times 2$ supercell containing 144 atoms was used to calculate the second-order interatomic force constants (IFCs) \cite{togo2015first} with a q-mesh of $3 \times 5 \times 3$ employed for phonon band structure calculation. The lattice thermal conductivity $\kappa_L$ was subsequently estimated using the modified Slack equation \cite{modifiedslackmodel} based on the calculated Debye temperature and Gr\"{u}neisen parameter. The thermodynamic stability was obtained by comparing the formation enthalpy of the target compound with its decomposed structures obtained from the experiment. A Python-based program named convex \cite{Utkarsh2020} was used to create phase diagrams via a convex hull approach \cite{Oses2018}.

The scattering rates and electronic transport properties were calculated using the AMSET code \cite{ganose2021efficient} that takes wave functions, band structure, deformation potential, mechanical and dielectric tensors as input. In AMSET, the scattering rates are obtained using Fermi's golden rule,
\begin{equation}
\tau_{i \rightarrow f}^{-1}
=
\frac{2\pi}{\hbar}
\left|
g_{fi}(\mathbf{k},\mathbf{q})
\right|^{2}
\delta\!\left(\varepsilon_i-\varepsilon_f\right),
\end{equation}

where \(i\) (\(f\)) denotes the initial (final) electronic state, \(\tau\) is the relaxation time, \(g_{fi}(\mathbf{k},\mathbf{q})\) represents the coupling matrix element between the initial and final states, and \(\varepsilon_i\) (\(\varepsilon_f\)) is the energy of the electron in the initial (final) state.
    
The matrices were constructed by including various scattering mechanisms, such as acoustic deformation potential (ADP), ionized impurity (IMP), and polar optical phonon (POP) scatterings. The total carrier scattering rate was then obtained by combining the contributions from the individual scattering mechanisms according to Matthiessen's rule,

\begin{equation}
\frac{1}{\tau_{\mathrm{tot}}}
=
\frac{1}{\tau_{\mathrm{ADP}}}
+
\frac{1}{\tau_{\mathrm{IMP}}}
+
\frac{1}{\tau_{\mathrm{POP}}},
\end{equation}
where \(\tau_{\mathrm{tot}}\) is the total relaxation time, while
\(\tau_{\mathrm{ADP}}\), \(\tau_{\mathrm{IMP}}\), and
\(\tau_{\mathrm{POP}}\) denote the relaxation times associated with
acoustic deformation potential (ADP), ionized impurity (IMP), and
polar optical phonon (POP) scattering mechanisms, respectively. 

The obtained relaxation time was used to calculate various transport properties by solving the Boltzmann Transport Equation in AMSET. Various transport coefficients, including  seebeck coefficient (\(S = \frac{1}{qT}\frac{\mathcal{L}^{1}}{\mathcal{L}^{0}}\)), electrical conductivity (\(\sigma = \mathcal{L}^{0}\)), and electronic thermal conductivity (\(\kappa_e = \frac{1}{q^{2}T}\left[\frac{(\mathcal{L}^{1})^{2}}{\mathcal{L}^{0}} - \mathcal{L}^{2}\right]\)) are obtained from the generalized transport equation,
\begin{equation}
\mathcal{L}^{\alpha}(\mu,T)
=
q^{2}
\int
\Sigma(\varepsilon)
(\varepsilon-\mu)^{\alpha}
\left(
-\frac{\partial f^{0}(\varepsilon,T)}
{\partial \varepsilon}
\right)
\, d\varepsilon ,
\end{equation}
where \(q\) is the electronic charge, \(\Sigma(\varepsilon)\) is the spectral conductivity, \(\mu\) is the chemical potential, and \(f^{0}(\varepsilon,T)\) is the equilibrium Fermi-Dirac distribution function \cite{ganose2021efficient,madsen2018boltztrap2,onsager1931reciprocal}.

Next, the optical properties were evaluated by the \textit{}$G_{0}W_{0}$ approximation based on Many Body Perturbation Theory (MBPT) \cite{hedin1965new,shishkin2006implementation}. A k-point grid of $10\times 18 \times 10$ was employed to calculate the KS energies. For all optical response calculations, a $3\times 2\times 3$ k-point grid was used with a total of 144 bands, which include both occupied and unoccupied states. For better accuracy, the number of bands considered were approximately 4 to 5 times the number of occupied bands. The inclusion of \textit{e-h} interaction plays a significant role in calculating the optical properties, so we incorporated the Bethe-Salpeter-Equation (BSE) on QP energies \cite{salpeter1951relativistic,hanke1979many,sander2015beyond}. To construct the BSE Hamiltonian in the basis of valence and conduction bands, 12 occupied and the first 32 unoccupied bands were considered after a convergence check for the KMgX compounds.

\section{Result and Discussion} \label{sec:result}
\subsection{Phases of KMgX}
\subsubsection{Structure Analysis}	
The KMgX (X = P, As, Sb, Bi) family is experimentally known to crystallize in tetragonal structure with space group $P4/nmm$ (No. 129) \cite{vogel1979neue}. This structure exhibits a layered arrangement along the c-direction, consisting of a K-layers sandwiched between the Mg–X layers, where Mg and X atoms form edge-sharing tetrahedra (see Fig.~\ref{fig:str}a). Owing to the electropositive nature of K, charge transfer from K to the MgX framework results in an ionic interaction between K$^{+}$ and [MgX]$^{-}$ layers. 

In addition to the tetragonal phase, KMgX compounds have been reported in two hexagonal polymorphs: a centrosymmetric $P6_{3}/mmc$ phase (see Fig.~\ref{fig:str}b) and polar non-centrosymmetric $P6_{3}mc$ phase (see Fig.~\ref{fig:str}c) \cite{bennett2013orthorhombic,bennett2012hexagonal}. Both phases feature a layered structure consisting of Mg-X layers separated by K-layers. The Mg–X layer form a planar hexagonal framework in the $P6_{3}/mmc$ space group (see Fig.~\ref{fig:str}b), whereas they adopt a buckled hexagonal framework in the $P6_{3}mc$ space group (see Fig.~\ref{fig:str}c). The polar ${P6_{3}mc}$ phase can be derived directly from the non-polar ${P6_{3}/mmc}$ phase through a translationengleiche (t) subgroup transition involving loss of inversion symmetry and preserving the hexagonal framework.

\begin{table*}[t]
	\centering
	\renewcommand{\arraystretch}{1.4} 
	\setlength{\tabcolsep}{8.8pt}      
	\caption{Optimized lattice parameters (in \AA) of KMgX (where X = P, As, Sb, and Bi) compounds in the various considered phases, calculated within the GGA framework.}
	
	\begin{tabular}{c cc cc cc c ccc cc}
		
		\hline\hline
		
		\multirow{2}{*}{\textbf{System}} 
		& \multicolumn{2}{c}{\textbf{Tetragonal} \cite{vogel1979neue}} 
		& \multicolumn{2}{c}{\textbf{Hexagonal}} 
		& \multicolumn{2}{c}{\textbf{Hexagonal}} 
		& \textbf{Cubic} 
		& \multicolumn{3}{c}{\textbf{Orthorhombic}}
		& \multicolumn{2}{c}{\textbf{Expt.}} \\
		
		& \multicolumn{2}{c}{\textbf{($P4/\mathrm{nmm}$)}} 
		& \multicolumn{2}{c}{\textbf{($P6_{3}/\mathrm{mmc}$)}} 
		& \multicolumn{2}{c}{\textbf{($P6_{3}\mathrm{mc}$)}} 
		& \textbf{($F\bar{4}3m$)}  
		& \multicolumn{3}{c}{\textbf{($Pnma$)}} 
		& \multicolumn{2}{c}{\textbf{($P4/\mathrm{nmm}$)}} \\
		
		\cline{2-13}
		
		& \textbf{a} & \textbf{c/a} 
		& \textbf{a} & \textbf{c/a} 
		& \textbf{a} & \textbf{c/a} 
		& \textbf{a} 
		& \textbf{a} & \textbf{b/a} & \textbf{c/a} 
		& \textbf{a} & \textbf{c/a} \\
		
		\hline
		
		\textbf{KMgP}  & 4.463 & 1.713 & 4.368 & 2.339 & -- & -- & 6.847 & 8.055 & 0.564 & 0.961 & 4.446 & 1.697 \\
		
		\textbf{KMgAs} & 4.576 & 1.720 & 4.511 & 2.293 & -- & -- & 7.032 & 8.247 & 0.566 & 0.972 & 4.546 & 1.697 \\
		
		\textbf{KMgSb} & 4.838 & 1.726 & 4.849 & 2.164 & 5.083 & 1.606 & 7.443 & 8.688 & 0.571 & 0.985 & 4.812 & 1.704 \\
		
		\textbf{KMgBi} & 4.933 & 1.734 & 4.972 & 2.135 & 5.210 & 1.599 & 7.609 & 8.856 & 0.574 & 0.996 & 4.881 & 1.717 \\
		
		\hline\hline
		
	\end{tabular}
	\label{tab:lattice_parameters_kmgx}
\end{table*}

A highly symmetric cubic structure with space group $F\bar{4}3m$ has also been theoretically proposed for KMgX compounds (Fig.~\ref{fig:str}d). However, this zinc-blende derived phase has not yet been experimentally realized. But several first-principles studies have explored its electronic, elastic, and thermoelectric properties \cite{chand2025enhanced,lv2025unconventional,arif2016elastic}. Unlike the layered tetragonal and hexagonal phases, the cubic structure exhibits a three-dimensional network in which the K-atoms occupy the interstitial sites within the Mg-X network (shown in the Fig.~\ref{fig:str}d), resulting in a highly symmetric crystal arrangement. 
	
We propose a previously unexplored orthorhombic phase with space group $Pnma$, obtained as a translationengleiche subgroup of the parent tetragonal $P4/nmm$ structure. On transition from tetragonal to orthorhombic structure, the layered Mg-X network in the $a-b$ plane gets distorted into an interconnected three-dimensional framework through symmetry lowering (see in Fig.~\ref{fig:str}e). Figure~\ref{fig:str} depicts that the tetragonal and hexagonal phases retain their structural characteristics, while the cubic and the orthorhombic phases lose the layered characteristic by converting to a 3D framework. The symmetry-driven structural evolution from the tetragonal to the orthorhombic phase alters the electronic, vibrational, transport, and photovoltaic properties of the orthorhombic KMgX compounds.
    
To investigate the relative stability of the tetragonal, cubic, hexagonal, and orthorhombic phases of KMgX compounds, their energy-volume data are fitted with the third order Birch--Murnaghan equation of state as shown in Fig.~\ref{fig:eos}. The optimized structural parameters of all KMgX compounds are summarized in Table~\ref{tab:lattice_parameters_kmgx}. The tetragonal phase is experimentally established as the ground-state structure of KMgX compounds, the calculated lattice parameters are benchmarked against available experimental data \cite{vogel1979neue}. The optimized values are slightly overestimated, with a deviation of less than 1.4\% from the experimental data. From KMgP to KMgBi, a systematic expansion of the lattice is observed in all structural phases, consistent with the increasing atomic radius of the pnictogen atoms.

 \subsubsection{Phase Transition}\label{sec:Pt}
    	\begin{figure}
    	\centering
    	\includegraphics[width=0.45\textwidth]{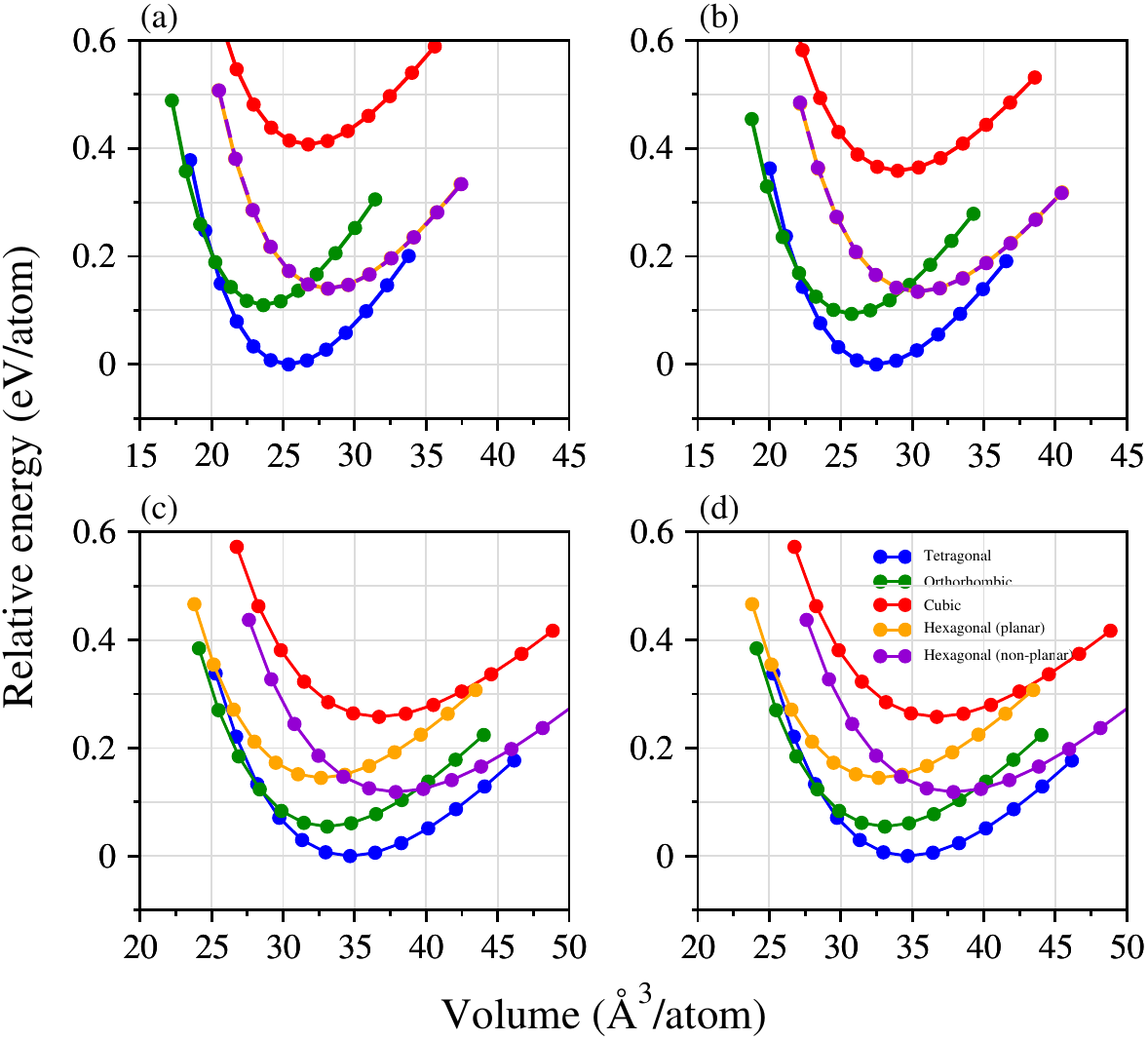}
    	\caption{The third-order Birch-Murnaghan fitted equation of state (EOS) plots, showing relative energy (eV/atom) as a function of volume per atom (\AA$^{3}$/atom) for KMgX compounds (X = P, As, Sb, and Bi) in cubic ($F\bar{4}3m$), hexagonal ($P6_{3}mc$), hexagonal ($P6_{3}/mmc$), tetragonal ($P4/nmm$), and orthorhombic ($Pnma$) symmetries. Panels (a), (b), (c), and (d) correspond to KMgP, KMgAs, KMgSb, and KMgBi, respectively. The ground-state tetragonal (${P4/nmm}$) symmetry in each case is set as zero on the vertical axis, represented by a blue line.}
    	\label{fig:eos}
\end{figure}
The calculated equation-of-state (EOS) in Fig.~\ref{fig:eos} confirms the tetragonal structure (blue curve) as the ground-state phase of all KMgX compounds, in agreement with experimental observations \cite{vogel1979neue}. Beyond the tetragonal phase, the orthorhombic phase emerges as the lowest energy metastable phase, as shown in the Fig.~\ref{fig:eos}. The energy–volume curves of the tetragonal and orthorhombic phases intersect for all KMgX compounds, suggesting pressure driven structural transformation between these two phases. However, several other metastable phases, such as hexagonal and cubic, remain energetically less favorable over the investigated volume range. In case of KMgP and KMgAs, the EOS curves correspond to non-polar $P6_{3}/mmc$ and polar $P6_{3}mc$ structure overlap, indicating the absence of a stable polar distortion, whereas for KMgSb and KMgBi, both polar and non-polar hexagonal phases are retained as distinct stable structure \cite{bennett2013orthorhombic}. 

\begin{table*}
	\centering
	\caption{Calculated band gaps ($E_g$) and the transition pressures for the tetragonal to orthorhombic phase transition in the KMgX (X = P, As, Sb, Bi) compounds. The symbols D and I in parentheses denote direct and indirect nature of the band gaps, respectively. The superscript $a$ represents the experimentally observed phase, where $b$ indicates the inclusion of spin-orbit coupling.}
	\renewcommand{\arraystretch}{1.4} 
	\setlength{\tabcolsep}{8pt}      
	
	\begin{tabular}{c cccc cccc c}
		\hline\hline
		
		& \multicolumn{8}{c}{Band Gap $E_g$ (eV)} & Transition Pressure \\
		
		\cline{2-9}
		
		Compound & \multicolumn{4}{c}{Tetragonal$^{a}$ \cite{chaudhary2023effect}} 
		& \multicolumn{4}{c}{Orthorhombic} 
		& $(P_t$ in GPa) \\
		
		\cline{2-9}
		
		& PBE & PBE$^{b}$ & HSE06 & HSE06$^{b}$ 
		& PBE & PBE$^{b}$ & HSE06 & HSE06$^{b}$ & \\
		
		\hline
		
		KMgP  & 1.80 (I) & 2.02 (I) & 2.57 (I) & 2.72 (I) 
		& 1.68 (D) & 1.67 (D) & 2.43 (D) & 2.41 (D) & 12.01 \\
		
		KMgAs & 1.24 (I) & 1.37 (I) & 1.95 (I) & 2.06 (I) 
		& 1.10 (D) & 0.93 (D) & 1.69 (D) & 1.60 (D) & 10.32 \\
		
		KMgSb & 1.29 (I) & 1.24 (I) & 2.04 (I) & 1.85 (I) 
		& 1.10 (D) & 0.93 (D) & 1.77 (D) & 1.58 (D) & 6.02 \\
		
		KMgBi & 0.49 (I) & 0.16 (I) & 1.10 (I) & 0.62 (I) 
		& 0.05 (D) & 0.02 (D) & 0.50 (D) & 0.003 (D) & 6.27 \\
		
		\hline\hline
		
	\end{tabular}
	\label{table:band_gap}
\end{table*}
\begin{figure*}
	\centering
	\includegraphics[width=0.90\textwidth]{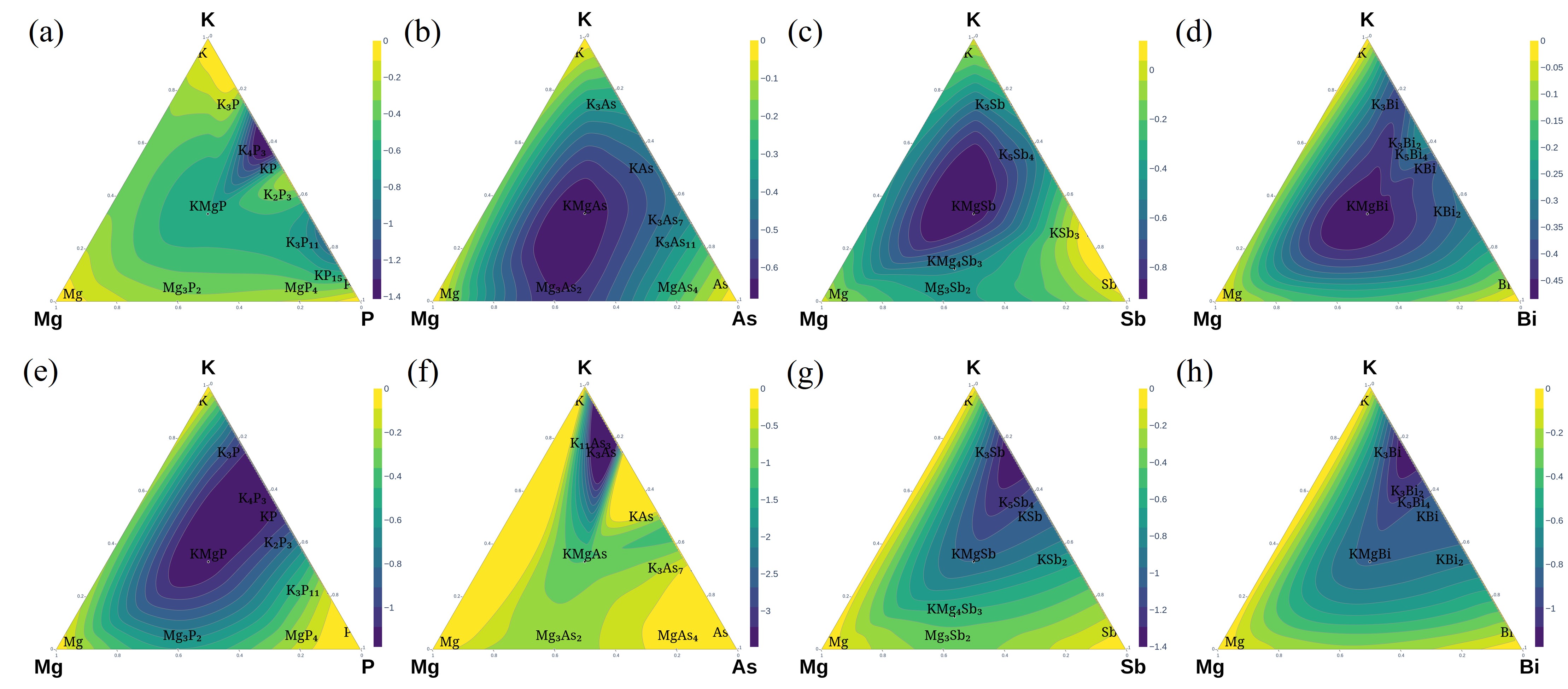}
	\caption{Ternary phase diagrams for KMgX (X = P, As, Sb, Bi) compounds, with color contours representing the formation enthalpy (eV/atom). Panels (a)--(d) correspond to the tetragonal $P4/nmm$ structures at ambient pressure, while (e)--(h) correspond to the orthorhombic $Pnma$ structures under external pressure.}
	\label{fig:td}
\end{figure*}

Although KMgX has been computationally investigated in multiple phases, the Fig.~\ref{fig:eos} reveals that the orthorhombic phase is energetically closest to the tetragonal ground state. This observation suggests that the tetragonal to orthorhombic transformation is the most likely structural transition pathway. Such a phase transition may be driven by external stimuli, particularly hydrostatic pressure, which can modify the relative phase stability through volume reduction.

To estimate the transition pressure associated with the tetragonal to orthorhombic transformation, the common-tangent method is applied to the energy–volume (E–V) curves of the competing phases. Thermodynamically, the pressure is defined as the negative slope of the E–V curve, $P = -\frac{dE}{dV}$. Accordingly, the slope of the common tangent yields the equilibrium transition pressure at which the two phases coexist. The calculated transition pressures for all KMgX (X= P, As, Sb, and Bi) compounds are summarized in Table~\ref{table:band_gap}. 

\subsection{\label{sec:3B}KMgX – Orthorhombic}
The predicted orthorhombic phase arises from a pressure-induced phase transformation from the parent tetragonal structure. While this phase has not yet been experimentally reported for the KMgX family, its existence is supported by numerous experimental studies on related ternary compounds that crystallize in the same $Pnma$ symmetry. For example, MgPtSi has been experimentally synthesized in the orthorhombic TiNiSi-type ($Pnma$) structure \cite{kudo2015superconductivity}, where distortion of a high symmetry framework drives the formation of a lower-symmetry orthorhombic phase. Similarly, several XYZ-type ternary silicides, including CoVSi, CoNbSi, and CoTaSi, have also been reported as stable orthorhombic phases \cite{singh2020first}. These observations suggest that the emergence of an orthorhombic phase in KMgX compounds is structurally plausible under appropriate thermodynamic conditions.
\begin{figure*}
	\centering
	\includegraphics[width=0.90\textwidth]{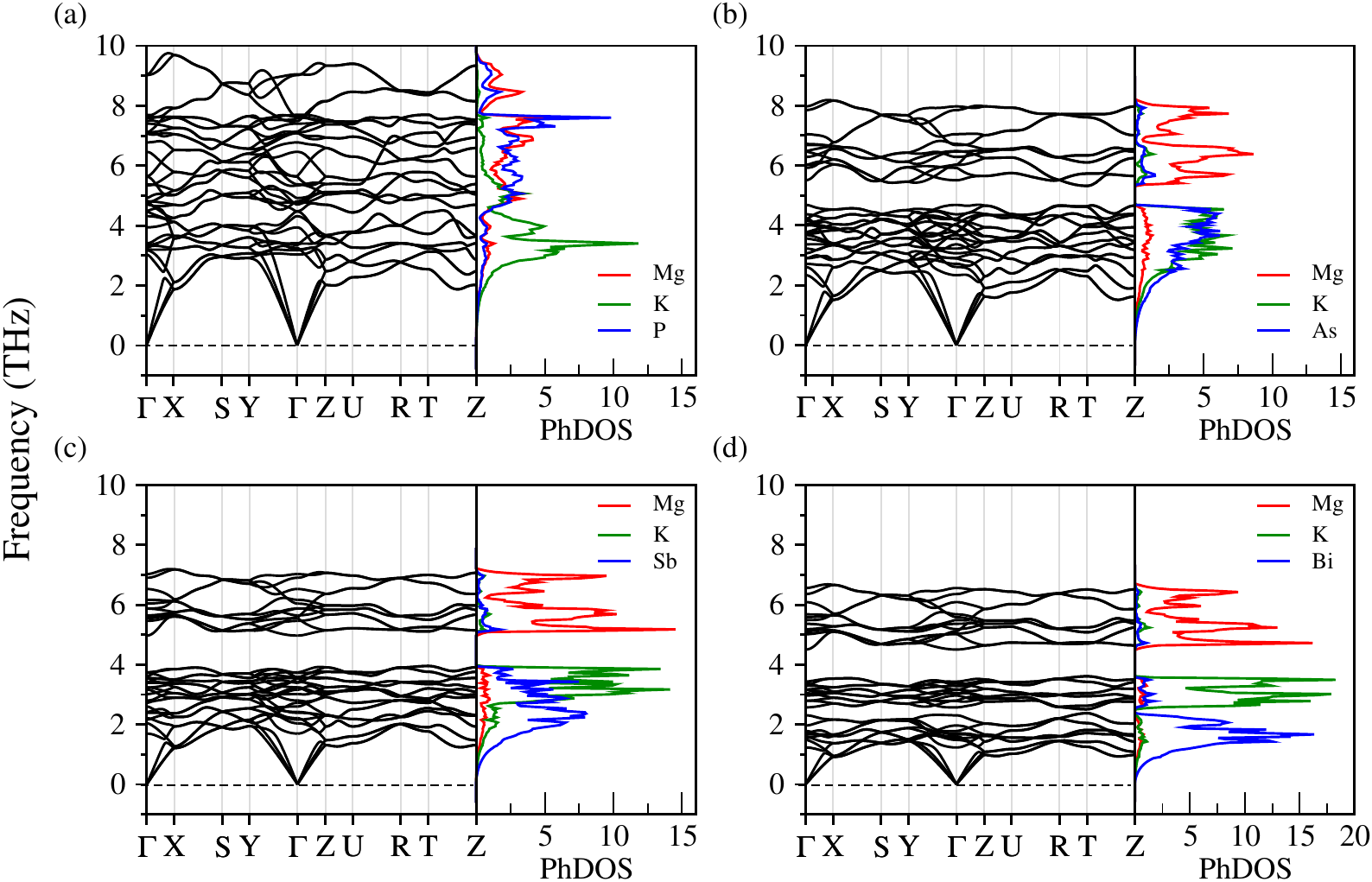}
	\caption{Phonon dispersion relations and corresponding phonon density of states, PhDOS (in states THz$^{-1}$) of the (a) KMgP, (b) KMgAs, (c) KMgSb, and (d) KMgBi compounds in orthorhombic ($Pnma$) phase.}
	\label{fig:ph}
\end{figure*}

Motivated by these studies, we examined the stability of the proposed orthorhombic phase. For that, we assessed the thermodynamic stability by carefully analysing the total enthalpies of all the competing phases and decompositions. We systematically calculate the per-atom change in formation enthalpy of all KMgX compounds at ambient conditions and above the transition pressure. The formation enthalpy per atom ($\Delta H_f^{\mathrm{atom}}$) of the proposed orthorhombic phase of the KMgX compounds was computed relative to their constituent elemental phases using the following expression
\begin{equation*}
            \Delta H_f^{\mathrm{atom}} =\frac{
H_{\mathrm{comp}}(P) - \sum_i n_i H_i(P)
}{
\sum_i n_i
}
\end{equation*}
where $H_{\mathrm{comp}}(P)$ is the enthalpy of the compound at 'P' pressure, $H_i(P)$ is the enthalpy of the $i^{\mathrm{th}}$ element of the compound at 'P' pressure, and $n_i$ is the number of atoms of the $i^{\mathrm{th}}$ element in the compound. More negative formation enthalpies indicate enhanced stability relative to competing phases and decomposition products.

The thermodynamic stability of the KMgX compounds was further evaluated through convex-hull analysis, (Fig.~\ref{fig:td}), where the color contouring helps to visualize the location of the hull minima, which indicate the stability of the phases. The formation enthalpies calculated at ambient pressure and above the transition pressure reveal a clear evolution in phase stability. At ambient conditions, the tetragonal phase occupies the convex-hull minimum and therefore remains the thermodynamic ground state. In contrast, upon exceeding the transition pressure, the orthorhombic phase becomes the thermodynamically favorable phase, indicating pressure-induced stabilization.
	
After establishing thermodynamic stability, we next examine dynamic, and mechanical stability. The static stability of the proposed orthorhombic phase is already confirmed in section ~\ref{sec:Pt} from the equation of state (EOS) analysis. The dynamic stability of the orthorhombic phase is examined through the lattice-dynamical calculations. The phonon dispersions are presented in Fig.~\ref{fig:ph}. Across all KMgX compounds, no imaginary phonon frequencies are observed throughout the entire Brillouin zone. This confirms the dynamic stability of the orthorhombic phase of each KMgX compound. 
	 
Figure~\ref{fig:ph} depicts a systematic evolution of phonon spectra across the series from KMgP to KMgBi. The maximum phonon frequency decreases from $\sim$10 THz in KMgP to about $\sim$7 THz in KMgBi, reflecting the increasing atomic mass of the pnictogen species. This trend is consistent with the harmonic relation, $\omega \propto \sqrt{\frac{k}{m}}$, where k is the effective bond force constant, and m is the reduced atomic mass. Another notable feature is the emergence of a gap between the optical modes for the heavier compounds, which may have important consequences for phonon transport and lattice thermal conductivity.
	
Next, the mechanical stability of the orthorhombic phase is examined by calculating its elastic constants using the Density Functional Perturbation Theory (DFPT) method. For orthorhombic symmetry, nine independent elastic constants are obtained, and the complete elastic tensor matrix is provided in the Supplementary Information. All KMgX compounds satisfy the Born mechanical stability criteria (see Supplementary Information), confirming their mechanical robustness in the $Pnma$ phase. Additional parameters, including Young’s modulus, shear modulus, ductility, Poisson’s ratio, and Debye temperature, are summarized in Table S1 of the Supplementary Information. For all KMgX compounds, the bulk modulus remains below 37~GPa, indicating relatively soft and compressible structure. Such mechanical softness supports the feasibility of inducing the orthorhombic phase through the application of external pressure.
	
Collectively, the proposed orthorhombic KMgX phase satisfies all relevant stability criteria, including thermodynamic, static, dynamic, and mechanical. These results demonstrate that the $Pnma$ structure is not merely a hypothetical configuration but rather a viable metastable phase that can be stabilized under suitable conditions. The established stability provides a strong foundation to investigate the properties across various domains.
	

\subsection{\label{sec:3a}Electronic properties}
\begin{figure}[h]
	\centering
	\includegraphics[width=0.460\textwidth]{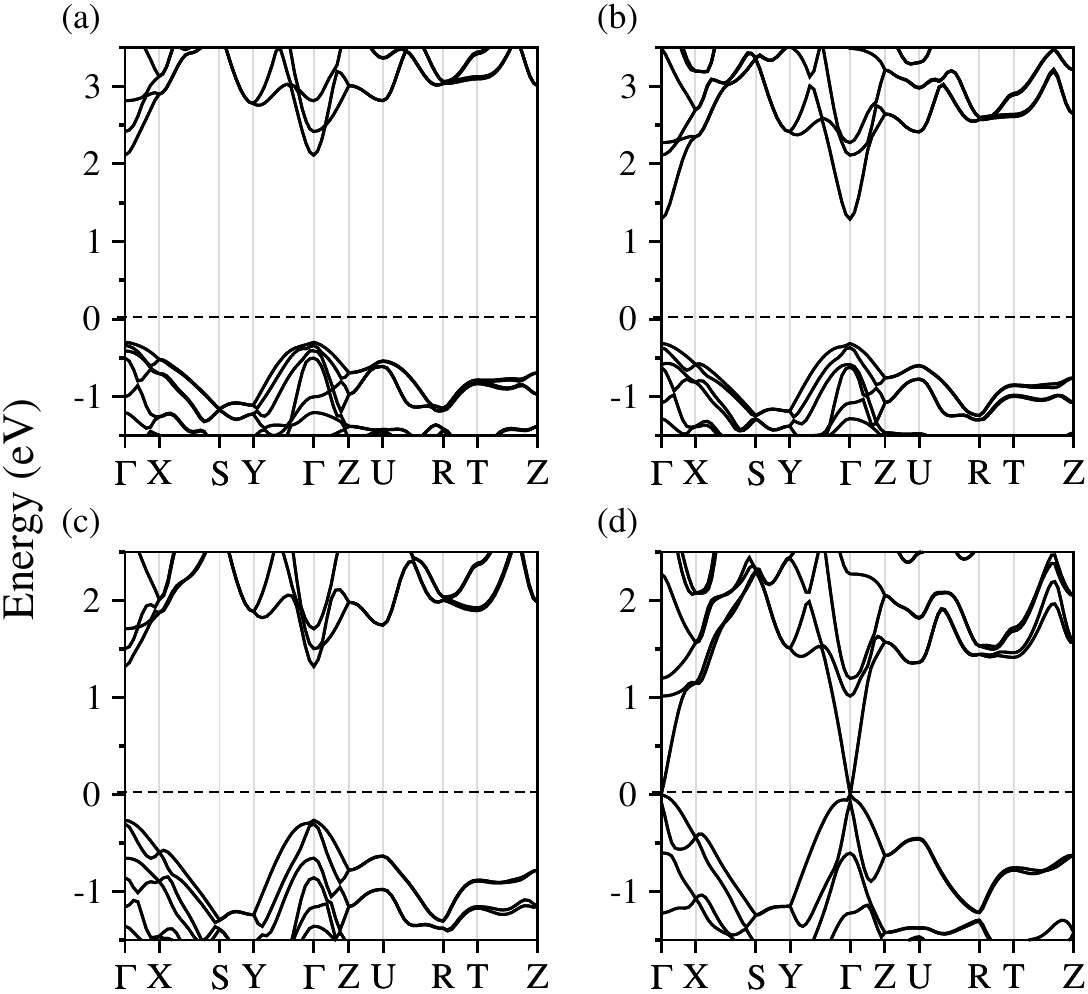}
	\caption{The Kohn-Sham (KS) band structure using the hybrid functional (HSE06) with spin-orbit coupling (SOC) for (a) KMgP, (b) KMgAs, (c) KMgSb, and (d) KMgBi compounds in orthorhombic ($Pnma$) symmetry.}
	\label{fig:bs}
\end{figure}

The electronic band structures of orthorhombic KMgX (X = P, As, Sb, and Bi) compounds, calculated using the HSE06 hybrid functional with spin–orbit coupling (SOC), are shown in Fig.~\ref{fig:bs}. A systematic reduction in the band gap is observed from KMgP to KMgBi, which is attributed to the increase in atomic size of the pnictogen atom. As evident from the band-gap values listed in Table~\ref{table:band_gap}, the inclusion of SOC has only a marginal effect on KMgP and KMgAs owing to the relatively weak spin–orbit interaction of the lighter pnictogens. In contrast, KMgSb and KMgBi exhibit noticeable SOC-induced modifications, leading to a significant reduction in the band gap because of the stronger spin–orbit interaction associated with the heavier Sb and Bi atoms. These results highlight the increasing importance of spin-orbit coupling (SOC) across the KMgX series. A similar band-gap reduction with increasing atomic size and the enhanced effect of SOC for the heavier pnictogens has been reported previously for the other phase of KMgX compounds \cite{chaudhary2023effect}. 

KMgP, KMgAs, and KMgSb exhibit a direct band gap at the $\Gamma$ point in visible spectral range, highlighting their potential for optoelectronic and energy-conversion applications. In contrast, orthorhombic KMgBi shows semi-metallic behavior and anomalous electronic features in the presence of SOC. This behavior is reminiscent of the reported tetragonal phase of KMgBi, which has attracted considerable attention as a topological material \cite{le2017three}. The pronounced SOC effects observed in orthorhombic KMgBi suggest the possibility of nontrivial topological states. However, a detailed investigation of its topological properties is beyond the scope of the present work. The electronic properties relevant to photovoltaic and thermoelectric applications depend on the magnitude and nature of the band gap, so further discussions focus primarily on KMgP, KMgAs, KMgSb, and orthorhombic KMgBi is excluded from further analyses.

 \begin{figure*}
 	\centering
 	\includegraphics[width=0.95\textwidth]{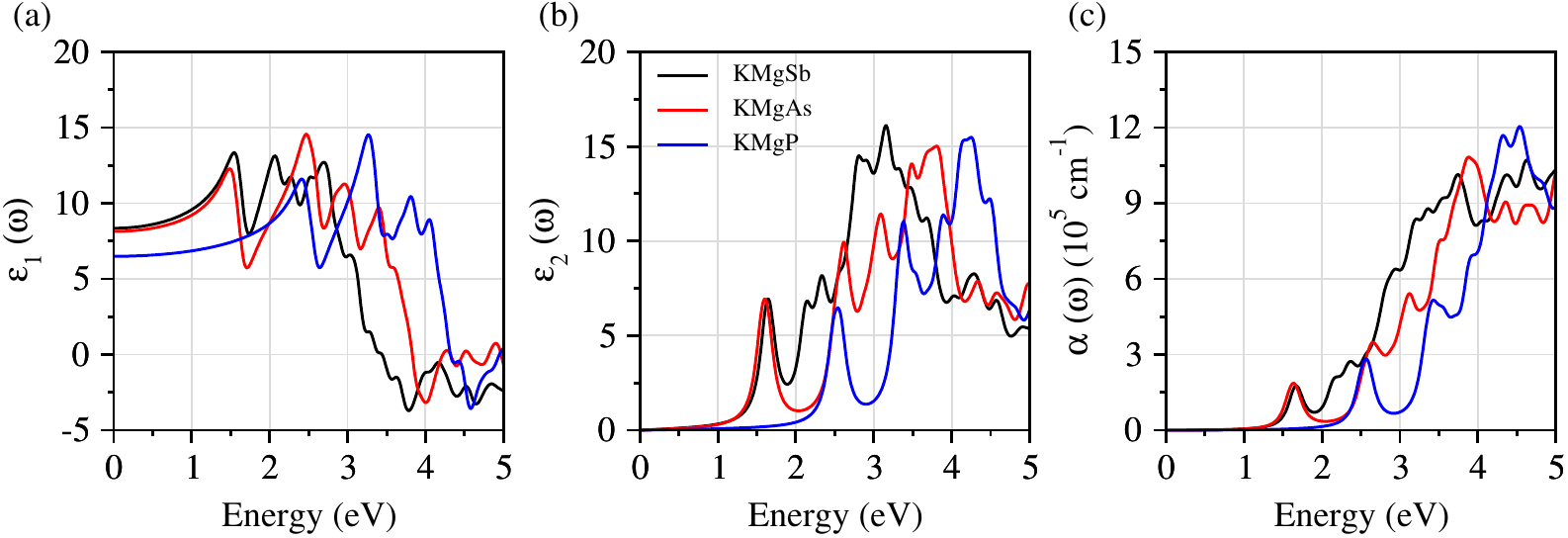}
 	\caption{The (a) real component $\varepsilon_1(\omega)$, (b) imaginary component $\varepsilon_2(\omega)$ of the dielectric function, and (c) absorption coefficient $\alpha(\omega)$ for the KMgX compounds ($X = \mathrm{P, As, Sb}$), calculated using the Bethe--Salpeter equation (BSE) based on quasiparticle energies obtained within the $G_0W_0$ approximation.}
 	\label{fig:dc}
 \end{figure*}

\subsection{\label{sec:3d}Optical Properties}
The electronic band structures demonstrate that the orthorhombic KMgP, KMgAs, and KMgSb compounds possess direct band gaps, suggesting their potential for optoelectronics and photovoltaic applications. Their direct band gaps enable efficient optical transitions and strong absorption of solar radiation without requiring phonon-assisted transitions, resulting in efficient photo-carrier generation. An important quantity for understanding the optical response of a material is the complex dielectric function, \( \varepsilon(\omega) \), i.e.,
\begin{equation}
    	 \varepsilon(\omega) = \varepsilon_1(\omega) + i\,\varepsilon_2(\omega)
\end{equation}
where the real component of the dielectric function (\( \varepsilon_1(\omega) \)) describes the dispersion of light inside the material, and the imaginary component of the dielectric function (\( \varepsilon_2(\omega) \)) describes the material's ability to absorb light. \( \varepsilon_1(\omega) \)  is subsequently derived from \( \varepsilon_2(\omega) \) through the Kramers--Kronig relation \cite{ambrosch2006linear}.
\begin{equation*}
		\varepsilon_1(\omega)=1+\frac{2}{\pi} \, P \int_{0}^{\infty}
		\frac{\omega' \varepsilon_2(\omega')}
		{\omega'^2-\omega^2}\, d\omega'
\end{equation*}
where P is the Cauchy principal value of the integral.
 From the dielectric function, other optical quantities, including real and imaginary parts of optical conductivity $\sigma(\omega)$, complex refractive index $n(\omega)$, absorption coefficient $\alpha(\omega)$, and reflectivity $r(\omega)$, can be easily computed \cite{ambrosch2006linear,kumar2022first}. To accurately capture many-body and excitonic effects, all optical properties are evaluated using the $G_0W_0$ approximation combined with the Bethe–Salpeter equation (BSE) framework. The BSE is solved within the Tamm--Dancoff approximation \cite{vorwerk2019bethe} that neglects the coupling between the (resonant) excitation and (anti-resonant) de-excitation terms.
	
Figure~\ref{fig:dc}a shows the real part of the dielectric function for the orthorhombic KMgX compounds. The calculated static dielectric constants are 6.49, 8.14, and 8.35 for KMgP, KMgAs, and KMgSb, respectively. A systematic enhancement in dielectric constant is observed from P to Sb, reflecting the increasing polarizability of the heavier pnictogen atoms. The enhanced dielectric screening lowers exciton binding energies, thereby facilitating electron-hole charge separation and carrier transport. Furthermore, all compounds exhibit pronounced dielectric response within the visible spectral range, indicating efficient interaction with solar radiation.
	
The imaginary part of the dielectric function, shown in Fig.~\ref{fig:dc}b, exhibits prominent excitonic absorption peaks at 2.54, 1.60, and 1.64~eV for KMgP, KMgAs, and KMgSb, respectively. These peaks appear close to the fundamental band gaps, a characteristic signature of direct-gap semiconductors with strong optical transitions. Further, the peak intensities indicate efficient photon absorption near the absorption onset. The optical absorption ($\alpha(\omega)$) of a material is governed by both the joint density of states (JDOS) and the optical transition strength, ($|\langle n | \hat{H} | m \rangle|^{2}$). 
 \begin{equation}
\alpha(\omega)=
\frac{2\pi}{\hbar}
\frac{2}{8\pi^{3}}
\int
\left|
\left\langle n \left| \hat{H} \right| m \right\rangle
\right|^{2}
\delta\left(
E_m(\vec{k})-E_n(\vec{k})-\hbar\omega
\right)
d^{3}k
\end{equation}
where $\hbar\omega$ is the photon energy, the integration spans over the entire Brillouin zone, and $\left\langle n \left| \hat{H} \right| m \right\rangle$ is the transition matrix element between the bands n and m.
To elucidate the origin of the strong optical response, we further analyzed the optical transition matrix elements along the high-symmetry points and identified a pronounced transition dipole moment at the $\Gamma$ point (see Fig. S16. in the Supporting Information), particularly along the crystallographic \textit{z}-direction. This behavior reveals significant optical anisotropy in the KMgX compounds. Since all compounds exhibit a direct band gap at the $\Gamma$ point, the large transition dipole moment at this k-point enhances the optical oscillator strength, confirming the presence of dipole-allowed optical transitions. 

The calculated absorption spectra further support this interpretation (see Fig.~\ref{fig:dc}c). The absorption coefficient associated with the first excitonic peak appears $\sim 10^{5}$~cm$^{-1}$ for all compounds, indicating strong optical absorption in the visible region. Among the series, KMgSb exhibits the largest absorption coefficient, reaching $2.83 \times 10^{5}$~cm$^{-1}$ near the absorption edge. This value is approximately one order of magnitude larger than that reported for the benchmark photovoltaic absorber CH$_3$NH$_3$PbI$_3$ ($3.8 \times 10^{4}$~cm$^{-1}$ at 2.0~eV) \cite{shirayama2016}, suggesting that KMgSb possesses favorable light-absorption characteristics for thin-film photovoltaic applications.
	
\begin{table}[ht]
		\centering
		\caption{Calculated band gap ($E_g$) obtained from G$_0$W$_0$ calculations and the photovoltaic performance parameters of KMgX (X = P, As, and Sb) compounds predicted by the spectroscopic limited maximum efficiency (SLME). Here, $J_{sc}$ represents the short-circuit current density and $V_{oc}$ denotes the open-circuit voltage. The symbol D in parentheses indicates the direct nature of the band gap.}
		\renewcommand{\arraystretch}{1.4} 
		\setlength{\tabcolsep}{3.5pt}      
		
\begin{tabular}{c c c c c c}
		\hline\hline
			
			\textbf{Compositions} 
			& \textbf{$E_g$ (eV)} 
			& \textbf{$\alpha(\omega)$} 
			& \textbf{$J_{sc}$} 
			& \textbf{$V_{oc}$} 
			& \textbf{SLME} \\
			
			& \textbf{($G_0W_0$)} 
			& \textbf{($cm^{-1}$)} 
			& \textbf{(mA/cm$^2$)} 
			& \textbf{(V)} 
			& \textbf{(\%)} \\
			
			\hline
			
			KMgP  & 2.79 (D) & $1.76 \times 10^{5}$ & 3.18  & 2.40 & 7.19  \\
			
			KMgAs & 1.77 (D) & $1.86 \times 10^{5}$ & 20.46 & 1.45 & 27.12 \\
			
			KMgSb & 1.81 (D) & $2.83 \times 10^{5}$ & 19.36 & 1.49 & 26.40 \\
			
			\hline\hline
		\end{tabular}
		\label{tab:slme}
\end{table}
	
To assess the photovoltaic potential of these materials, we employed the spectroscopic limited maximum efficiency (SLME) \cite{yu2012identification} as a screening parameter. Unlike the conventional Shockley–Queisser (SQ) limit \cite{shockley1961detailed}, which estimates the maximum photovoltaic efficiency solely on the basis of the band gap, SLME incorporates the actual optical absorption characteristics, the nature of the band gap (direct or indirect), and the absorber thickness. Thus, SLME provides a more realistic assessment of the maximum achievable power conversion efficiency as compared to SQ limit. The calculated SLME values are summarized in Table~\ref{tab:slme}.
	
\begin{figure} 
		\centering
		\includegraphics[width=0.35\textwidth]{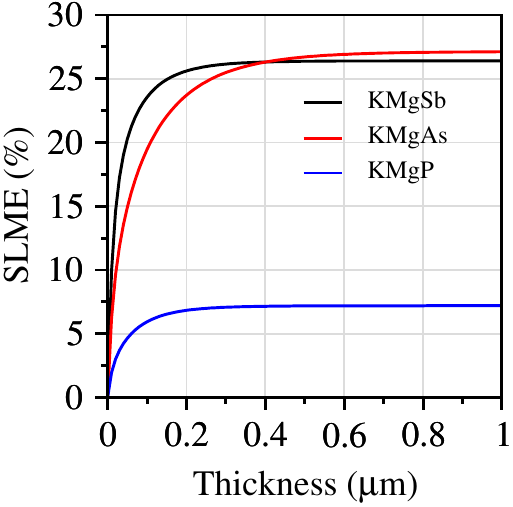}
		\caption{Comparison of the spectroscopic limited maximum efficiency (SLME) of all KMgX (X = P, As, and Sb) compounds, obtained using the absorption coefficient from the Bethe-Salpeter equation.}
		\label{fig:slme}
\end{figure}
	
Figure~\ref{fig:slme} shows the thickness-dependent SLME for orthorhombic KMgP, KMgAs, and KMgSb. For all compounds, the SLME increases rapidly with film thickness up to approximately 0.2~$\mu\mathrm{m}$, followed by a gradual saturation beyond 0.6~$\mu\mathrm{m}$. This behavior reflects the efficient absorption of incident photons within relatively thin absorber layers. A clear compositional trend is observed across the series, with photovoltaic performance improving from KMgP to KMgSb. The enhancement correlates with the reduced exciton binding energy and stronger dielectric screening of the heavier pnictogen compounds.
	
At a thickness of 0.3~$\mu\mathrm{m}$, KMgP, KMgAs, and KMgSb exhibit SLME values of 7.079\%, 25.488\%, and 26.149\%, respectively. Notably, the SLME values of KMgAs and KMgSb are significantly higher than that reported for GaAs (15\%) at a thickness of 0.4 $\mu\mathrm{m}$ \cite{yin2015superior,solet2025many}. However, the maximum predicted SLME for GaAs reaches approximately 28\% at a thickness of 3~$\mu\mathrm{m}$, which is slightly higher than those of orthorhombic KMgAs and KMgSb. Below 1~$\mu\mathrm{m}$, KMgAs and KMgSb can be considered promising alternatives to GaAs-based thin film solar cells.

The maximum SLME further increase to 27.12\% for KMgAs and 26.40\% for KMgSb. These values compare favorably with several widely studied photovoltaic absorbers, including $\mathrm{CH_3NH_3PbI_3}$ (24.94\%) \cite{shirayama2016} and $\mathrm{CsPbI_3}$ (22.32\%) \cite{Mei2025}. These results identify orthorhombic KMgAs and KMgSb as promising candidates for next-generation thin-film photovoltaic devices. Finally, we compared the photovoltaic performance of the newly proposed orthorhombic phase with that of the experimentally known tetragonal phase using literature-reported SLME values (see Fig. S17. in the Supplementary Information). The orthorhombic phase consistently exhibits superior photovoltaic performance across the KMgX series. This enhancement demonstrates that the pressure-induced tetragonal-to-orthorhombic transformation not only alters the crystal structure but also substantially improves the optoelectronic characteristics relevant to solar-energy conversion. The combination of strong optical absorption, favorable dielectric screening, and high predicted conversion efficiencies establishes orthorhombic KMgAs and KMgSb as attractive photovoltaic materials.

\subsection{\label{sec:3f}Thermoelectric Properties}
The thermoelectric (TE) performance of a material is quantified by the dimensionless figure of merit, (zT),
\begin{equation}
	zT=\frac{S^{2}\sigma T}{\kappa}
	=\frac{S^{2}\sigma T}{\kappa_{\mathrm{e}}+\kappa_{\mathrm{L}}}.
\end{equation}
where (S) is the Seebeck coefficient, ($\sigma$) is the electrical conductivity, and ($\kappa$) is the total thermal conductivity comprising electronic ($\kappa_{\mathrm{e}}$) and lattice ($\kappa_{\mathrm{L}}$) contributions. For a good thermoelectric material, a high zT achieve by simultaneous enhancement of the power factor ($S^{2}\sigma$) and suppression of total thermal conductivity ($\kappa$).

S, $\sigma$, and $\kappa_{\mathrm{e}}$ are the electronic transport coefficients calculated within the rigid-band approximation, in which different doping concentrations are simulated by shifting the Fermi level while keeping the underlying electronic band structure unchanged. 
All transport calculations are performed using the HSE06-corrected electronic band structures, as hybrid functionals generally provide a more reliable description of the band gap. Consistent with the electronic structure analysis presented in section~\ref{sec:3a}, SOC has negligible influence on the band structures of KMgP and KMgAs near the Fermi level; therefore, their transport properties are calculated without SOC. In contrast, the transport properties of KMgSb are evaluated by considering the spin-orbit coupling.

Carrier transport properties of KMgX compounds are calculated by considering acoustic deformation potential (ADP), ionized impurity (IMP), and polar optical phonon (POP) scattering mechanisms. The piezoelectric scattering is excluded because orthorhombic KMgX crystallizes in the centrosymmetric $Pnma$ structure. Among the considered scatterings, POP scattering dominates throughout the KMgX series and gives the carrier relaxation times in femtoseconds, ($10^{-15}$~s).

The carrier-concentration dependence of the Seebeck coefficient (S), electrical conductivity ($\sigma$), and electronic thermal conductivity ($\kappa_{\mathrm{e}}$) at different temperatures is presented for all KMgX compounds (see Fig. {S9-S11} of the Supplementary Information). At low carrier concentrations, all compounds exhibit relatively large Seebeck coefficients, which is characteristic of semiconducting systems. Among all, KMgSb exhibits the highest value of Seebeck coefficient ($501.75~\mu\mathrm{VK^{-1}}$) at 300~K under p-type doping with a carrier concentration of $10^{18}~\mathrm{cm^{-3}}$. In addition, the maximum value of electrical conductivity  reaches  $(0.7-4.9)\times10^{5}~\mathrm{S\,m^{-1}}$ over the investigated carrier-concentrations. As the temperature increases, the Seebeck coefficient increases, whereas the electrical conductivity is reduced owing to enhanced carrier scattering (see Fig. S12. in the Supplementary Information). The coexistence of comparatively large Seebeck coefficients and appreciable electrical conductivity results in favorable power factors across the series. 

\begin{figure} 
	\centering
	\includegraphics[width=0.48\textwidth]{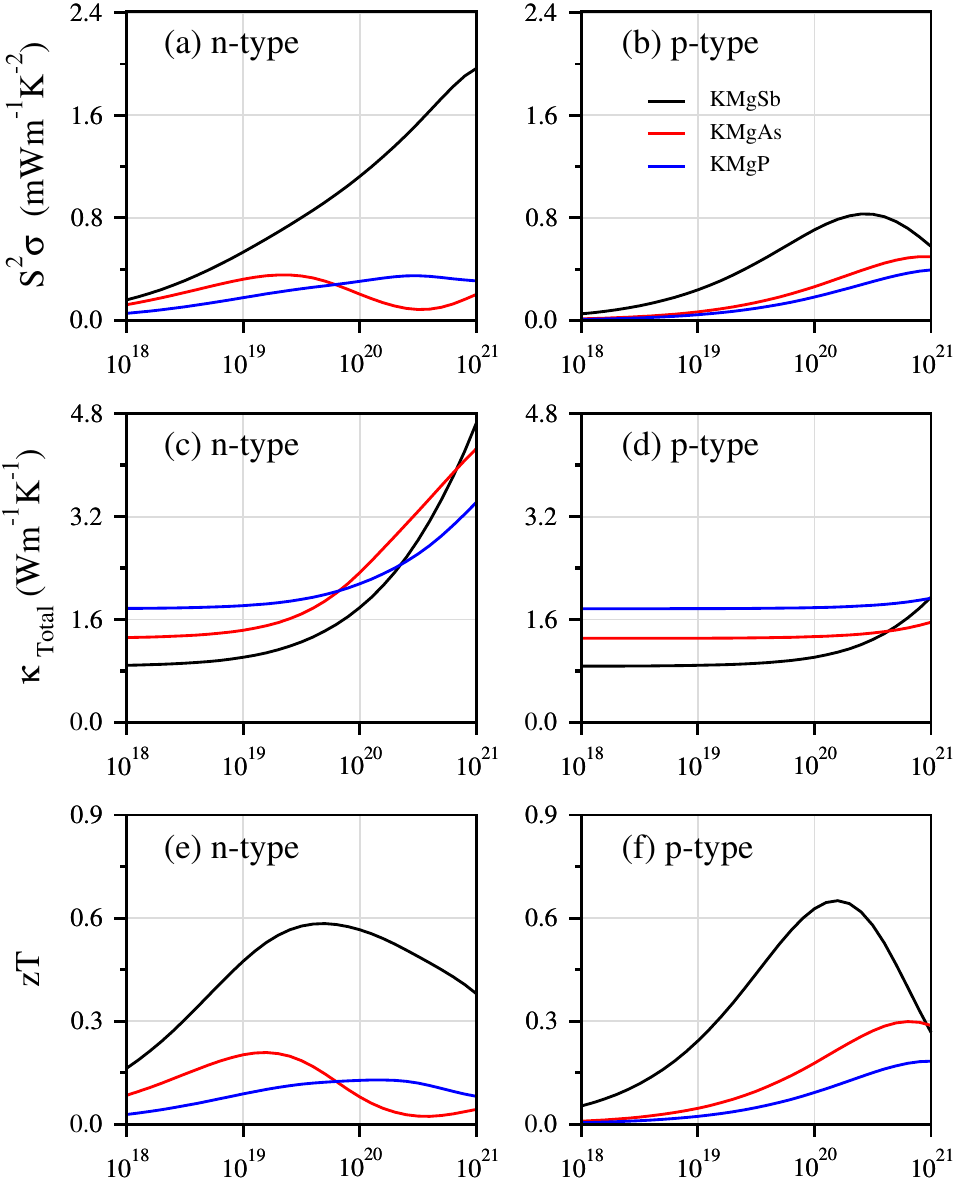}
	\caption{Calculated (a,b) power factor (PF), (c,d) total thermal conductivity ($\kappa_{\mathrm{tot}}$=$\kappa_e$+$\kappa_L$), and (e,f) thermoelectric figure of merit (zT) for the KMgX (X = P, As, and Sb) compounds at 900~K. The black, red, and blue lines represent KMgSb, KMgAs, and KMgP, respectively.}
	\label{fig:trans}
\end{figure}

The calculated power factors for all KMgX compounds at 900~K are shown in Fig.~\ref{fig:trans}. For both n-type and p-type carrier concentrations, the power factor increases from KMgP to KMgSb. The systematic enhancement in the power factor can be directly correlated with the evolution of the electronic band structure across the series. The progressive reduction in the band gap and increased band dispersion promote improved carrier transport, resulting in an overall enhancement of the power factor across the KMgX series. As shown in Fig.~\ref{fig:trans}, orthorhombic KMgSb exhibits the highest power factor of 1.96~$\mathrm{mW,m^{-1},K^{-2}}$ under n-type doping at 900~K, slightly exceeding the reported value of 1.82~$\mathrm{mW,m^{-1},K^{-2}}$ for tetragonal KMgSb under the same conditions \cite{chaudhary2023effect}.

The heat transport in a semiconductor is predominantly carried by lattice vibrations. Consequently, the lattice thermal conductivity ($\kappa_{\mathrm{L}}$) plays a critical role in determining the overall thermoelectric efficiency. Within the framework of the phonon Boltzmann transport equation, ($\kappa_{\mathrm{L}}$) can be expressed as 
		
\begin{equation}
		\kappa_{\mathrm{L}}=\frac{1}{NV}\sum_{\lambda} C_{\lambda}v_{\lambda}^{2}\tau_{\lambda},
\end{equation}	
where $C_{\lambda}$, $v_{\lambda}$, and $\tau_{\lambda}$ represent the mode-specific heat capacity, phonon group velocity, and phonon lifetime, respectively. This relationship highlights that lattice thermal transport is governed primarily by the propagation and scattering characteristics of phonons. To elucidate the microscopic origin of the thermal transport behavior in KMgX (X = P, As, Sb), we first analyze the phonon group velocities and Gr\"{u}neisen parameters, both of which are key descriptors of phonon-mediated heat transport. The phonon group velocity,
	
\begin{equation}
		v_{\lambda}=\frac{\partial\omega_{\lambda}}{\partial q},
\end{equation}
quantifies the rate at which vibrational energy propagates through the lattice, whereas the Gr\"{u}neisen parameter,
	
\begin{equation}
	\gamma_{\lambda}
	=
	-\frac{V}{\omega_{\lambda}}
	\frac{\partial\omega_{\lambda}}{\partial V}
\end{equation}
provides a measure of lattice anharmonicity and the strength of phonon-phonon interactions. In general, reduced phonon group velocities and enhanced anharmonicity lead to stronger phonon scattering, shorter phonon lifetimes, and consequently lower lattice thermal conductivity.
	
\begin{figure} 
		\centering
		\includegraphics[width=0.35\textwidth]{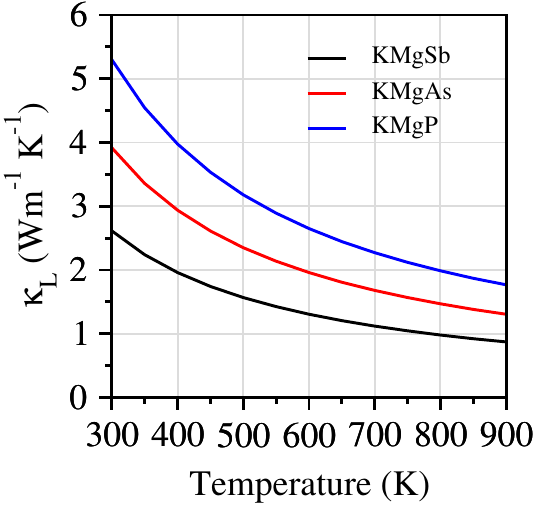}
		\caption{Temperature dependence of the calculated lattice thermal conductivity ($\kappa_L$) of the KMgX compounds. Black, red, and blue lines represent KMgSb, KMgAs, and KMgP, respectively.
		}
		\label{fig:KL}
\end{figure}
	
The calculated averaged Gr\"{u}neisen parameters for the KMgX compounds lie within the range of 1.41–1.46, as shown in Fig. S8. of the Supplementary Information, indicative of moderate intrinsic anharmonicity. The major contribution to the Gr\"{u}neisen parameter originates from the acoustic phonon branches constitutes the dominant heat-carrying modes for the KMgX compound. The strong anharmonicity of the acoustic phonons and relatively low phonon group velocities (as shown in Fig. S6. of the Supplementary Information) contribute significantly to the reduced lattice thermal conductivity.

\begin{table*}
	\centering
	\caption{The calculated lattice thermal conductivity ($\kappa_L$), Seebeck coefficient ($S$), electrical conductivity ($\sigma$), electronic thermal conductivity ($\kappa_e$), and thermoelectric figure of merit ($zT$) for KMgX (X = P, As, and Sb) compounds at 900 K under n-type and p-type doping conditions.}
	
	\renewcommand{\arraystretch}{1.4}
	\setlength{\tabcolsep}{5.8pt}
	
	\begin{tabular}{c c cccc cccc}
		
		\hline\hline
		
		& & \multicolumn{4}{c}{n-type} & \multicolumn{4}{c}{p-type} \\
		
		\cline{3-6}\cline{7-10}
		
		Composition &
		$\kappa_L$ &
		\multicolumn{1}{c}{$S$} &
		\multicolumn{1}{c}{$\sigma$} &
		\multicolumn{1}{c}{$\kappa_e$} &
		\multicolumn{1}{c}{$zT$} &
		\multicolumn{1}{c}{$S$} &
		\multicolumn{1}{c}{$\sigma$} &
		\multicolumn{1}{c}{$\kappa_e$} &
		\multicolumn{1}{c}{$zT$} \\
		
		&
		(W m$^{-1}$ K$^{-1}$) &
		($\mu$V K$^{-1}$) &
		($10^{4}$ S m$^{-1}$) &
		(W m$^{-1}$ K$^{-1}$) &
		&
		($\mu$V K$^{-1}$) &
		($10^{4}$ S m$^{-1}$) &
		(W m$^{-1}$ K$^{-1}$) &
		\\
		
		\hline
		
		KMgP  & 1.767 & -107.99 & 2.59 & 0.543 & 0.129 & 178.47 & 1.18 & 0.164 & 0.183 \\
		
		KMgAs & 1.306 & -171.02 & 1.18 & 0.198 & 0.208 & 197.81 & 1.21 & 0.163 & 0.299 \\
		
		KMgSb & 0.871 & -187.66 & 2.29 & 0.549 & 0.584 & 216.99 & 1.64 & 0.220 & 0.651 \\
		
		\hline\hline
		
	\end{tabular}
	
	\label{tab:thermoelectric}

\end{table*}

The lattice thermal conductivity ($\kappa_L$) can be evaluated using a variety of theoretical and empirical approaches. Among the available analytical models, the Slack equation \cite{modifiedslackmodel} is one of the most widely used methods for estimating the lattice thermal conductivity of crystalline solids. The temperature-dependent lattice thermal conductivities of the KMgX compounds are presented in Fig.~\ref{fig:KL}. For all compounds, ($\kappa_{\mathrm{L}}$) decreases with increasing temperature, reflecting the enhanced phonon–phonon Umklapp scattering at elevated temperatures. Furthermore, a systematic reduction in lattice thermal conductivity is observed with increasing pnictogen atomic size from KMgP to KMgSb.

The observed trend is also consistent with the phonon dispersion characteristics discussed in Section~\ref{sec:3B}. In particular, KMgAs and KMgSb exhibit significant overlap between the acoustic and low-lying optical phonon branches, resulting in the disappearance of a distinct acoustic–optical phonon gap. Such a feature substantially enlarges the three-phonon scattering processes involving acoustic and optical phonons, leading to a pronounced reduction in lattice thermal conductivity. 
The lattice thermal conductivity of KMgSb (2.64~$\mathrm{Wm^{-1}K^{-1}}$) evaluated using Slack model is lower than the reported value for its tetragonal phase (3.30~$\mathrm{Wm^{-1}K^{-1}}$) \cite{chaudhary2023effect}, which is obtained from a direct solution of the phonon Boltzmann transport equation. 

It should be noted that the Slack model provides an approximate description of lattice thermal conductivity based on averaged phonon properties and does not explicitly account for the full spectrum of phonon scattering processes. Consequently, the calculated values are often overestimated compared to those obtained from a rigorous solution of the phonon Boltzmann transport equation \cite{peter2025ultra}. As a consistency check, we applied the same Slack-model approach to the cubic and tetragonal phases of KMgSb. The calculated lattice thermal conductivities are 3.58 $\mathrm{Wm^{-1}K^{-1}}$ for the tetragonal phase and 1.42 $\mathrm{Wm^{-1}K^{-1}}$ for the cubic phase, compared with the corresponding reported values of 3.30 $\mathrm{Wm^{-1}K^{-1}}$ \cite{chaudhary2023effect} and 0.83 $\mathrm{Wm^{-1}K^{-1}}$ \cite{lv2025unconventional} respectively. In both phases, the Slack-model estimates exceed the reported values, indicating a systematic overestimation of $\kappa_{\mathrm{L}}$ in KMgSb. To establish the lower bound of thermal transport in the KMgX compounds, the minimum lattice thermal conductivity is further estimated using the Cahill model \cite{cahill1992lower}. The calculated minimum lattice thermal conductivities are 0.849, 0.558, and 0.474 $\mathrm{Wm^{-1}K^{-1}}$ for KMgP, KMgAs, and KMgSb, respectively. The minimum lattice thermal conductivity represents the amorphous limit of heat transport, which is transferred through highly localized vibrational modes. Therefore, the Cahill limit provides an important benchmark for assessing the ultimate potential for reducing lattice thermal conductivity.

The temperature-dependent thermoelectric figure of merit zT for the KMgX compounds is presented in Fig.~\ref{fig:trans}. A systematic enhancement in zT is observed across the series from KMgP to KMgSb, reflecting the combined effects of improved electronic transport and intrinsically low lattice thermal conductivity. Among the investigated compounds, KMgSb exhibits the highest thermoelectric performance, attaining a maximum zT value of 0.65 and 0.58 at 900~K under p-type and n-type doping respectively. Notably, the zT values obtained for n-type and p-type carrier concentrations are comparable throughout the investigated temperature range, indicating nearly symmetric thermoelectric performance for both carrier types. Such balanced behavior is advantageous for practical thermoelectric module design, as it enables the same material system to serve as both n-type and p-type elements. The maximum zT values at 900~K are summarized in Table~\ref{tab:thermoelectric}. For n-type doping, the maximum zT values are 0.13, 0.21, and 0.58 for KMgP, KMgAs, and KMgSb, respectively, while p-type doping yields corresponding maximum values of 0.18, 0.30, and 0.65. Since the lattice thermal conductivity is likely overestimated, the calculated zT values may provide conservative estimates of the actual thermoelectric performance. Among the KMgX compounds, KMgSb exhibits the most favorable thermoelectric characteristics, combining a high power factor with a comparatively low lattice thermal conductivity, and consequently delivers the highest zT for both carrier types.

\section{Conclusion}
In summary, we have systematically investigated the pressure-induced orthorhombic phase of the KMgX (X = P, As, Sb, and Bi) family using first-principles calculations. Structural optimization and equation-of-state analyses indicate that this phase can be accessed from the ground-state tetragonal structure under external pressure. The viability of the orthorhombic phase was further supported by comprehensive stability assessments, including static, dynamical, mechanical, and thermodynamic analyses, all of which confirm its stability.

Electronic structure calculations reveal that KMgP, KMgAs, and KMgSb are direct-band-gap semiconductors, whereas orthorhombic KMgBi exhibits semimetallic behavior. Consequently, the subsequent optoelectronic and thermoelectric analyses focus on KMgP, KMgAs, and KMgSb. All three compounds exhibit strong optical absorption in the visible region, with absorption coefficients on the order of $10^{5}$~cm$^{-1}$. Among the investigated phases, KMgSb exhibits the highest absorption coefficient of $2.83 \times 10^{5}$~cm$^{-1}$ near the absorption onset. The photovoltaic performance of the KMgX compounds was evaluated using the spectroscopic limited maximum efficiency (SLME). KMgAs and KMgSb exhibit the highest SLME values of 27.12\% and 26.40\%, respectively, at an absorber thickness of 0.6~$\mu\mathrm{m}$. The orthorhombic phase of the KMgX family exhibits consistently higher SLME values, highlighting the improved photovoltaic performance of the pressure-induced orthorhombic phase. Furthermore, the thermoelectric properties of the KMgX family were evaluated through the dimensionless figure of merit (zT). At 900~K, KMgSb exhibits the highest zT values of 0.58 and 0.65 for n-type and p-type doping, respectively. These favorable zT values suggest that KMgSb is a promising material for the realization of both n-type and p-type thermoelectric elements.

Overall, the present work demonstrates that pressure-induced phase transformation provides an effective route to access previously unexplored orthorhombic phases of the KMgX family. The combination of structural stability, direct-gap semiconducting behavior, optically allowed band-edge transitions, favorable photovoltaic efficiencies, and promising thermoelectric transport characteristics highlights the multifunctional nature of these compounds. Collectively, these results underscore the potential of the orthorhombic KMgX phases for energy-conversion applications and provide guidance for future experimental investigations.
	
\section*{Acknowledgement}
This work utilized the Supercomputing Facility at the Indian Institute of Technology Roorkee, established under the National Supercomputing Mission (NSM), Government of India, and supported by the Centre for Development of Advanced Computing (C-DAC), Pune, India. Additional computational resources were provided by the Institute Computer Centre (ICC), Indian Institute of Technology Roorkee. Financial support from the Ministry of Education, Government of India, is also acknowledged.

 \bibliographystyle{apsrev4-2}
\bibliography{kmx}

\end{document}